\documentclass[preprint1]{aastex}
\usepackage{amsmath}
\usepackage{multirow}
\usepackage{longtable}
\usepackage{threeparttable}
\slugcomment{}

\newcommand{\be}{\begin{equation}}
\newcommand{\ee}{\end{equation}}
\newcommand{\Swift}{{\it Swift\/}}
\shorttitle{GRB Luminosity Function}
\shortauthors{Cui, Wu, Wei, et al.}
\shortauthors{}
\begin{document}
\title{The Optical Luminosity Function of Gamma-ray Bursts deduced from ROTSE-III Observations}
\author{X.~H. Cui\altaffilmark{1,2}, X.~F. Wu\altaffilmark{1,3,4}, J.~J. Wei\altaffilmark{1}, F. Yuan\altaffilmark{5}, W.~K. Zheng\altaffilmark{6}, E.~W. Liang\altaffilmark{2,7}
C.~W.~Akerlof\altaffilmark{8},
M.~C.~B.~Ashley\altaffilmark{9},
H.~A.~Flewelling\altaffilmark{10},
E.~G\"{o}\v{g}\"{u}\c{s}\altaffilmark{11},
T. G\"{u}ver\altaffilmark{12},
\"{U}.~K{\i}z{\i}lo\v{g}lu\altaffilmark{13},
T.~A.~McKay\altaffilmark{8},
S.~B.~Pandey\altaffilmark{14},
E.~S.~Rykoff,\altaffilmark{15},
W.~Rujopakarn\altaffilmark{16,20},
B.~E.~Schaefer\altaffilmark{17},
J.~C.~Wheeler\altaffilmark{18},
S.~A.~Yost\altaffilmark{19}
}
\altaffiltext{1}{Purple Mountain Observatory, Chinese Academy of Sciences, Nanjing 210008,
China; xhcui@bao.ac.cn, xfwu@pmo.ac.cn, jjwei@pmo.ac.cn.
}
\altaffiltext{2}{National Astronomical Observatories, Chinese Academy of Sciences,
             Beijing 100012, China.
}
\altaffiltext{3}{Chinese Center for Antarctic Astronomy, Nanjing 210008, China.
}
\altaffiltext{4}{Joint Center for Particle, Nuclear Physics and Cosmology, Nanjing
University-Purple Mountain Observatory, Nanjing 210008, China.
}
\altaffiltext{5}{Research School of Astronomy and Astrophysics, The Australian National University, Weston Creek, ACT 2611, Australia; fang.yuan@anu.edu.au
}
\altaffiltext{6}{Department of Astronomy, University of California, Berkeley, CA 94720-3411, USA; zwk@astro.berkeley.edu
}
\altaffiltext{7}{Department of Physics and GXU-NAOC Center for Astrophysics and Space Sciences, Guangxi University,
Nanning 530004, China; lew@gxu.edu.cn
}
\altaffiltext{8}{Department of Physics, University of Michigan, Ann Arbor, MI 48109, USA}
\altaffiltext{9}{School of Physics,
        University of New South Wales, Sydney, NSW 2052, Australia}
\altaffiltext{10}{Institute for Astronomy, University of Hawaii, 2680 Woodlawn Drive, Honolulu HI 96822}
\altaffiltext{11}{Sabanc{\i} University, Orhanl{\i}-Tuzla 34956 {\.I}stanbul, Turkey}
\altaffiltext{12}{Department of Astronomy and Space Sciences, Istanbul University Science Faculty, 34119 Istanbul, Turkey}
\altaffiltext{13}{Middle East Technical University, 06531 Ankara, Turkey}
\altaffiltext{14}{ARIES, Manora Peak, Nainital 263129, Uttarakhand, India}
\altaffiltext{15}{SLAC National Accelerator Laboratory, Menlo Park, CA 94025}
\altaffiltext{16}{Department of Physics, Faculty of Science, Chulalongkorn University, 254 Phayathai Road, Pathumwan, Bangkok 10330, Thailand}
\altaffiltext{17}{Department of Physics and Astronomy, Louisiana State
        University, Baton Rouge, LA 70803, USA}
\altaffiltext{18}{Department of Astronomy, University of Texas, Austin, TX
        78712, USA}
\altaffiltext{19}{Department of Physics, College of St. Benedict/St. John's
  University, Collegeville, MN 56321, USA}
  \altaffiltext{20}{Kavli Institute for the Physics and Mathematics of the Universe (WPI), Todai Institutes for Advanced Study, The University of Tokyo, Japan}

\begin{abstract}
We present the optical luminosity function (LF) of gamma-ray bursts (GRBs) estimated from a uniform sample of 58 GRBs from observations with the Robotic Optical Transient Search Experiment III (ROTSE-III). Our GRB sample is divided into two sub-samples: detected afterglows (18 GRBs), and those with upper limits (40 GRBs). The $R$ band fluxes 100s after the onset of the burst for these two sub-samples are derived. The optical LFs at 100s are fitted by assuming that the co-moving GRB rate traces the star-formation rate. The detection function of ROTSE-III is taken into account during the fitting of the optical LFs by using Monte Carlo simulations. We find that the cumulative distribution of optical emission at 100s is well-described with an exponential rise and power-law decay (ERPLD), broken power-law (BPL), and Schechter LFs. A single power-law (SPL) LF, on the other hand, is ruled out with high confidence.

\end{abstract}


\section{Introduction}
Gamma-ray bursts (GRBs) are the most luminous explosions in the Universe. Although highly transient, they afford a good laboratory to study astrophysics in extreme conditions. The prompt gamma-ray emission refers to the emission component detected by the gamma-ray detector and is commonly interpreted as emission from internal shocks (e.g., Rees \& M{\'e}sz{\'a}ros 1994, 2005; Daigne \& Mochkovitch 1998; Pe'er et al. 2006) or internal magnetic energy dissipation processes (e.g., Usov 1992; Giannios \& Spruit 2006; Zhang \& Yan 2011). The prompt emission is often followed by an afterglow that is the multi-wavelength radiation from the external shock produced by interactions between the ejecta from the fireball and the ambient medium (see, e.g., Zhang \& M{\'e}sz{\'a}ros 2004 and Gao et al. 2013 for a review).

The identification of the first GRB redshift by Metzger et al. (1997; GRB 970508) revealed the cosmological origin and the vast energy release from GRBs, which allows detection out to extreme distances, e.g., $z\sim8.2$ (Tanvir et al. 2009; Salvaterra et al. 2009) and possibly at redshifts as high as $\sim9.4$ (Cucchiara et al. 2011). However, the transient nature of GRBs means that only a limited number have spectroscopic redshifts. This motivates a search for correlations between GRB luminosity and various observable parameters in order to derive {\it pseudo-redshifts} for GRB events without spectroscopic redshifts (e.g., Norris et al. 2000; Lloyd-Ronning et al. 2002; Firmani et al. 2004; Kocevski \& Liang 2006; Schmidt 2009). There is evidence for potential luminosity evolution in the gamma-ray band (Salvaterra et al. 2012), but the luminosity function (LF) is strongly dependent upon the instrumental detection function, and this complicates the interpretation of the results.

Following the identification of the first optical counterpart of a GRB in 1997 February 28 (van Paradijs et al. 1997), many statistical studies of optical afterglow (OA) light curves have been carried out, resulting in our understanding of the general features of the light curve. Dai et al. (2009) compared the cumulative distributions of peak gamma-ray photon fluxes and showed that \Swift\ and BATSE samples come from the same parent population of bursts. Morphological studies of light curves based on statistical analyses of large samples indicate that there are several emission components in the optical afterglow (e.g., Liang \& Zhang 2006; Panaitescu \& Vestrand 2008, 2011; Kann et al. 2010; 2011). Two universal tracks of the late optical luminosity light curves have been found (e.g., Nardini et al. 2006, Kann et al. 2006). Kann et al. (2010, 2011) compared optical light curves of different types of GRBs in the pre-\Swift\ and \Swift\ eras to study the distribution of early luminosities at 43.2 seconds in the bursts' rest frame, with known redshifts and host-galaxy extinctions. They found that the luminosity distribution can be approximated by three Gaussians. Typical features of GRB OA light curves comprise an early bump and plateau components (Panaitescu \& Vestrand 2008, 2011; Li et al. 2012; Liang et al. 2013; Wang et al. 2013). Wang et al. (2013) found that a single power-law provides good description of the LF at 10$^3$s. Most of the optical data for these studies, however, were collected from inhomogeneous observations with different instruments. Another issue is that optical observations often only start after the end of prompt gamma-ray emission. The optical LF of GRB afterglows is therefore poorly known since no complete sample within a given threshold is available. The detection of an optical counterpart of a GRB depends on the instrument, exposure time, observation epoch, etc. Therefore, a homogeneous data set from a single instrument, e.g., the Robotic Optical Transient Search Experiment III (ROTSE-III) in this work, can reduce these uncertainties. An analysis of early (e.g., at 100s) and homogeneous data after the onset of the burst is desirable to facilitate the interpretation of GRBs' optical LF.

Although the \Swift\ satellite has led to an increase in the number of GRBs with good redshift determinations, the sample is still not sufficiently large to directly measure the LF, and is affected by various biases. Since long GRBs are associated with the death of massive stars, the assumption that GRB rate traces the star-formation rate (SFR) has been used by many studies to constrain the GRB LF (e.g., Lamb \& Reichart 2000; Choudhury \& Srianand 2002; Natarajan et al.\ 2005; Daigne et al.\ 2006). Following this methodology, and assuming the most recent SFR determinations, we derive the optical LF of GRBs by fitting the observed ROTSE-III flux distributions at 100s after the prompt gamma-ray emission. This allows us to construct the optical LF of GRBs without needing to know their redshifts. The assumption we make is that the global rate of GRB OAs is proportional to the SFR and LF.

In this paper, we take advantage of the large sample of GRBs observed by ROTSE-III to explore the shape of the optical LF at early emission phases. The paper is organized as follows. In section 2, we present our GRB sample and the method we used to reproduce the flux distribution of the GRBs; our results are described in section 3; and section 4 concludes with a discussion. In this paper, we adopt $\Omega_m=0.3$, $\Omega_\Lambda=0.7$, and $H_0=70$ km s$^{-1}$ Mpc$^{-1}$.

\section{Observational data and Methodology\label{sec:method}}

ROTSE-III is a network of four identical 0.45-m telescopes distributed around the world to promptly observe OA of GRBs (Akerlof et al. 2003). We use a sample of ROTSE-III data to derive the GRB optical LF. The observed GRB OA rate is assumed to be a convolution of the optical LF with the cosmic GRB rate history. Since the intrinsic LF shape is affected by the instrumental sensitivity, we use simulations to determine this effect.

\subsection{ROTSE-III Observations}

We selected our sample of GRBs from ROTSE-III observations between 2005 February and 2011 July. For uniformity, we defined an epoch for brightness measurement to be 100s after the burst (with an equivalent exposure time of 5s). A non-detection, namely a 3$\sigma$ upper limit measurement, is also considered if it meets the following two criteria. First, we only consider GRBs that were triggered by the \Swift\ satellite, in order to have a uniform solid angle of sky coverage (see Section \ref{sec:OpticalLF} below). Second, the GRB must have ROTSE-III observations both before and after 100s, thereby allowing an interpolation to 100s. We have not included GRB 080319B, the naked-eye GRB, which was observed under inclement conditions with CCD condensation (Swan et al. 2008). For upper-limit measurements, we used a transformation factor to allow for different exposure times. Since most of the upper-limit measurements were with an exposure of 5s, we normalized all the longer exposure times (either 20s or 60s) to 5s exposures. Some GRBs have optical detections in exposures longer than 5s, but their observed magnitudes are fainter than the equivalent 5s limiting magnitude of the instrument (e.g., GRB050401; 8 GRBs in total, which are marked with stars in Table 1). These bursts are considered to be non-detections for the purpose of optical LF construction.

Our final sample consists of 58 GRBs with 18 detections and 40 upper-limit measurements. Some of the detections have been published earlier (Yost et al., 2007a,b; Rykoff et al. 2009; Yuan et al. 2008; Yuan 2010). For unpublished data, we use the ROTSE-III photometry package (RPHOT; Rykoff et al. 2009) to perform PSF photometry. Since all the ROTSE-III observations were taken unfiltered, and the response of the instrument is approximately the $R_C$ band (Rykoff et al. 2009), we adopt $R_C$ as our bandpass for photometry.

We take into account and make corrections for extinction in both our Galaxy and the GRB's host galaxy. Galactic extinction ($A_{\rm V}$) is corrected for using the values given by Schlafly \& Finkbeiner (2011) as listed in Table (\ref{data}). We transform the value of $A_{\rm V}$ to $A_R$ by applying an average extinction law (Cardelli et al. 1989). GRB 110625A is located in a region with very high extinction, $A_{V}=30.29$ (Schlafly \& Finkbeiner 2011), making it difficult to give any constraints on the LF based on this burst, which was therefore excluded from our non-detection sample. For the host galaxy extinction, we adopt a mean value of $A_V=0.2$ (Kann et al. 2010). We did not consider the uncertainties involved in assuming a mean $A_V$, including the effects of redshift and the change of attenuation law, since it is difficult to quantify the exact values of these effects. Finally, we corrected the flux from the observer frame to the cosmological rest frame with $F(\nu,t)=\kappa F_{\rm obs}(\nu,t)$, where $F_{\rm obs}(\nu,t)$ is the flux in the observer frame. The parameter $\kappa$ is defined by $\kappa =(1+z)^{\beta_o-\alpha_o-1}$ (with the convention $F(\nu,t)\propto \nu^{-\beta_o} t^{-\alpha_o}$), where we adopted the spectral index $\beta_o=0.75$ and power-law index $\alpha_o=1$ for the light curves of the optical afterglows. We list basic properties of the GRBs, namely the start and end observed time ($t_{\rm start}$ \& $t_{\rm end}$), coordinates (R.A. \& Dec), Galactic extinction $A_{\rm V}$, and the observed flux $F_{\rm obs}$ of the GRBs in our samples comprising of 18 detections and 40 upper limits in Table \ref{data}.

\begin{center}
\begin{longtable}{lllllllll}
\caption{Properties of the ROTSE-III GRB sample at 100s after trigger} \label{data} \\
\hline
GRB &$t_{\rm start}$ &$t_{\rm end}$ &R.A.  & Dec   & $A_{\rm V}$ &$F_{\rm obs}$    \\
 &s&  s&  (J2000)  & (J2000)  & mag & erg cm$^{-2}$ s$^{-1}$\\%**\\
\hline
\endfirsthead
\\
\multicolumn{7}{c}{18 detected GRBs}\\
\\
\hline
050801	&	21.8	&	10357.0	&	13:36:34.6	&	-21:55:48.0	&	0.255	&	3.22	\\
051109A	&	35.4	&	14534.9	&	22:01:15.8	&	40:51:00.0	&	0.502	&	4.0	\\
051111	&	29.4	&	8561.0	&	23:12:32.6	&	18:22:01.2	&	0.426	&	7.51	\\
060605	&	49.4	&	6677.6	&	21:28:30.7	&	-06:04:15.6	&	0.137	&	5.30	\\
060729	&	64.5	&	3045.9	&	06:21:08.9	&	-62:13:15.6	&	0.146	&	4.74	\\
061007	&	27.2	&	15051.9	&	03:05:11.8	&	-50:29:45.6	&	0.054	&	15.40	\\
080413A	&	20.4	&	3190.9	&	19:09:12.2	&	-27:40:37.2	&	0.441	&	15.06	\\
080603B	&	23.0	&	18238.0	&	11:46:13.0	&	68:03:39.6	&	0.033	&	17.02	\\
080607	&	22.0	&	4792.3	&	12:59:51.4	&	15:54:36.0	&	0.060	&	1.49	\\
080703	&	33.6	&	3926.1	&	06:47:17.3	&	-63:12:39.6	&	0.192	&	3.91	\\
080804	&	19.6	&	8442.3	&	21:54:42.0	&	-53:11:20.4	&	0.043	&	1.83	\\
080810	&	35.3	&	9643.6	&	23:47:07.9	&	00:18:36.0	&	0.075	&	4.28	\\
081008	&	41.9	&	4450.9	&	18:39:52.3	&	-57:25:58.8	&	0.252	&	4.86	\\
081029	&	86.7	&	3911.4	&	23:07:06.2	&	-68:10:44.4	&	0.083	&	4.16	\\
090418A	&	19.0	&	519.9	&	17:57:16.8	&	33:24:25.2	&	0.116	&	3.87	\\
090530	&	17.4	&	2448.7	&	11:57:36.0	&	26:35:24.0	&	0.063	&	2.67	\\
090618	&	24.7	&	45834.9	&	19:36:01.9	&	78:21:07.2	&	0.231	&	18.12	\\
110213A	&	27.2	&	6806.5	&	02:51:54.7	&	49:16:40.8	&	0.865	&	10.47	\\
\hline
\\
\multicolumn{7}{c}{40 upper-limit GRBs}\\
\\
\hline
050215A	&	65.2	&	200.2	&	23:13:36.7	&	49:19:40.8	&	0.591	&	6.57	\\
050306	&	64.8	&	185.4	&	18:49:14.2	&	-09:09:07.2	&	1.855	&	38.68	\\
050401*	&	33.2	&	281.2	&	16:31:29.5	&	02:11:06.0	&	0.177	&	6.32	\\
050822	&	31.8	&	100.9	&	03:24:25.4	&	-46:01:48.0	&	0.041	&	8.75	\\
051001	&	85.7	&	191.9	&	23:23:56.2	&	-31:30:54.0	&	0.041	&	4.59	\\
060110*	&	27.0	&	400.1	&	04:50:56.9	&	28:25:40.8	&	1.666	&	29.10	\\
060111B*	&	32.8	&	728.6	&	19:05:49.4	&	70:22:48.0	&	0.297	&	12.95	\\
060116	&	79.0	&	2993.2	&	05:38:47.5	&	-05:26:16.8	&	0.697	&	7.66	\\
060614	&	26.8	&	189.2	&	21:23:30.5	&	-53:01:37.2	&	0.058	&	8.88	\\
060904B*	&	19.3	&	6608.3	&	03:52:52.3	&	-00:43:44.4	&	0.472	&	4.43	\\
060927*	&	16.8	&	1768.7	&	21:58:11.3	&	05:22:12.0	&	0.165	&	3.53	\\
061121*	&	21.7	&	1169.5	&	09:48:54.7	&	-13:11:16.8	&	0.121	&	14.13	\\
061222A	&	47.2	&	115.1	&	23:53:01.0	&	46:31:26.4	&	0.266	&	3.25	\\
070208	&	40.8	&	1004.3	&	13:11:33.8	&	61:56:31.2	&	0.041	&	5.52	\\
070419A	&	81.3	&	1829.2	&	12:11:01.2	&	39:54:10.8	&	0.075	&	5.69	\\
070429A	&	96.7	&	1069.9	&	19:50:46.8	&	-32:25:12.0	&	0.460	&	7.41	\\
070611	&	44.7	&	9904.5	&	00:08:01.0	&	-29:45:21.6	&	0.035	&	5.35	\\
070621	&	24.1	&	1786.7	&	21:35:13.4	&	-24:48:32.4	&	0.130	&	4.14	\\
070704	&	94.3	&	1186.6	&	23:38:49.7	&	66:15:25.2	&	4.888	&	2750	\\
070808	&	29.6	&	1839.4	&	00:27:02.6	&	01:10:48.0	&	0.068	&	3.92	\\
071001	&	50.7	&	1058.9	&	09:58:49.7	&	-59:45:46.8	&	2.356	&	106.7	\\
071025	&	80.0	&	2981.4	&	23:40:15.6	&	31:47:02.4	&	0.195	&	3.66	\\
071118	&	83.9	&	1001.2	&	19:59:21.4	&	70:07:48.0	&	0.945	&	8.01	\\
080229A	&	32.8	&	1856.8	&	15:12:52.8	&	-14:41:49.2	&	0.398	&	19.27	\\
080303	&	20.4	&	1826.6	&	07:28:04.6	&	-70:13:51.6	&	0.511	&	147.9	\\
080330*	&	22.3	&	20297.3	&	11:17:06.7	&	30:36:25.2	&	0.044	&	1.18	\\
080604	&	84.7	&	1833.3	&	15:47:50.4	&	20:33:25.2	&	0.130	&	3.45	\\
080903	&	24.7	&	2500.6	&	05:47:09.6	&	51:15:21.6	&	0.560	&	5.62	\\
080916A	&	26.3	&	1043.0	&	22:25:09.4	&	-57:01:33.6	&	0.051	&	32.08	\\
081121	&	57.1	&	3125.1	&	05:57:07.7	&	-60:36:43.2	&	0.135	&	204.6	\\
090407	&	40.8	&	1102.7	&	04:35:55.0	&	-12:41:02.4	&	0.180	&	18.96	\\
090621A	&	60.8	&	889.9	&	00:43:56.9	&	61:56:16.8	&	6.025	&	11370	\\
090709A	&	26.6	&	1259.7	&	19:19:46.6	&	60:43:40.8	&	0.242	&	4.19	\\
090904A	&	85.5	&	1076.3	&	06:43:25.2	&	50:14:06.0	&	0.262	&	6.17	\\
091208A	&	30.5	&	1262.4	&	00:01:10.8	&	65:40:48.0	&	4.247	&	461.6	\\
091221	&	25.8	&	4964.8	&	03:43:11.5	&	23:14:34.8	&	0.568	&	7.46	\\
100621A	&	33.0	&	1021.6	&	21:01:14.2	&	-51:06:07.2	&	0.082	&	3.62	\\
100802A	&	34.1	&	1776.2	&	00:09:55.7	&	47:45:07.2	&	0.322	&	14.94	\\
110315A	&	49.2	&	1323.1	&	18:36:49.2	&	17:32:13.2	&	0.669	&	42.96	\\
110726A*	&	14.0	&	429.0	&	19:06:51.1	&	56:04:12.0	&	0.206	&	10.87	\\
\hline
\hline
\end{longtable}
\begin{tablenotes}
  \item[*]* These bursts were moved to the upper-limit sample from the detected sample based on the 5s limiting magnitude of the instrument.
  %\item[**]** $10^{-12}$ erg cm$^{-2}$ s$^{-1}$
  \end{tablenotes}
\end{center}

\subsection{Optical Luminosity Functions\label{sec:OpticalLF}}
The observed rate of GRB OAs with peak fluxes between $F_1$ and $F_2$ is
\begin{equation}
\label{dNdt}
\frac{dN}{dt}(F_1<F<F_2)=\int ^{z_{\rm max}}_0 \int ^{L(F_2,z)}_{L(F_1,z)} \Phi(L)\frac{R_{\rm GRB}(z)}{1+z}\frac{\Delta \Omega}{4\pi}\frac{dV(z)}{dz}dLdz,
\end{equation}
where the factor $(1+z)^{-1}$ is a result of cosmological time dilation, and the parameter $\Delta \Omega$=1.4 sr is the solid angle covered on the sky by \Swift\ (Salvaterra \& Chincarini 2007; we only consider GRBs that are triggered by \Swift), and $dV(z)/dz$ is the co-moving
volume element. The co-moving GRB formation rate is assumed to trace the cosmic SFR as
\begin{equation}
\label{RGRB}
R_{\rm GRB}(z)=k R_{\rm SFR}(z),
\end{equation}
where the factor $k$ is a constant. The SFR, $R_{\rm SFR}(z)$, in units of M$_\odot$ Mpc$^{-3}$ yr$^{-1}$, is parameterized following Hopkins \& Beacom (2006) as
\begin{equation}
\label{SFR}
\log R_{\rm SFR}(z)=a+b \log(1+z),
\end{equation}
with
\begin{equation}\label{SFRab}
(a,b)=\left\{ \begin{array}{l@{\quad\quad } l}
(-1.70,3.30),    \ \ \ \ \ \ \ \   z<0.993 \\
(-0.727,0.0549),\ \ 0.993<z<3.80  \\
(2.35,-4.46),  \ \ \ \ \ \ \ \ z>3.80\\
                \end{array}
\right.
.
\end{equation}
The maximum redshift $z_{\rm max}$ is determined by the Lyman $\alpha$ absorption of the emission in R band.

In this work, we compare the beaming-convolved LF of GRBs $\Phi(L)$ with four model functions:\\
(1) a single power law (SPL):
\begin{equation}
\label{L1}
\Phi(L)=\frac{1}{L_*}\left(\frac{L}{L_*}\right)^{\alpha_L};
\end{equation}
(2) a broken power law (BPL):
\begin{equation}
\label{L2}
\Phi(L)=\frac{1}{L_*}\left[\left(\frac{L}{L_*}\right)^{\alpha_{L1}}+\left(\frac{L}{L_*}\right)^{\alpha_{L2}}\right]^{-1};
\end{equation}
(3) an exponential rise and power-law decay function (ERPLD):
\begin{equation}
\label{L3}
\Phi(L)=\frac{1}{L_*}\left(\frac{L}{L_*}\right)^{\alpha_L}\exp\left(-\frac{L_*}{L}\right);
\end{equation}
(4) and a Schechter function:
\begin{equation}
\label{L4}
\Phi(L)=\frac{1}{L_*}\left(\frac{L}{L_*}\right)^{\alpha_L}\exp\left(-\frac{L}{L_*}\right),
\end{equation}
where $\alpha_L$ and $L_*$ are parameters determined by fitting the observational data.

The observed rate of GRB OAs is governed by the LF $\Phi(L)$ and the GRB formation rate $R_{\rm GRB}(z)$ based on fitted parameters including the factor $k$, $\alpha_{L}$ and $L_{\ast}$. The constant $k$ can be removed by normalizing the cumulative flux distribution of GRBs to $N(F_{\rm min}, F_{\rm max})$ as
\begin{equation}
\mathcal{N}(<F)=\frac{N(F_{\rm min},F)}{N(F_{\rm min}, F_{\rm max})}.
\end{equation}
We search for the best model parameters by evaluating the consistency between the cumulative flux distribution of the observed and expected GRBs with the one-sample Kolmogorov-Smirnov (K-S) test. In this test, the maximum value of the absolute difference between two cumulative distribution functions, D-stat, is evaluated with a significance level Prob. A larger value of Prob indicates a better consistency. A value of $\rm Prob>0.1$ is generally acceptable to claim statistical consistency, while a value of $\rm Prob<10^{-4}$ rejects the hypothesis of the consistency at high confidence.

\subsection{ROTSE-III Sensitivity Function}

In order to correct our observed LF for instrumental effects, we performed a simulation based on the number count distribution of the 40 GRBs in our upper-limit sample to reconstruct the detection function (i.e., the sensitivity function) of ROTSE-III. The simulation is a four-step process as follows. First, we construct a histogram of the flux limit from the 40 GRB limits. Second, a smoothed broken power-law (SBPL) is used to fit this histogram in the observed flux interval,
\begin{equation}
\label{SBPL}
N=N_0\left[\left(\frac{f}{f_{\rm b}}\right)^{\omega {\alpha_1}}+\left(\frac{f}{f_{\rm
b}}\right)^{\omega {\alpha_2}}\right]^{-1/\omega},
\end{equation}
where the parameter $N_0$ is a normalization factor, the parameter $f_{\rm b}$ is the flux at the break point of the SBPL, parameters $\alpha_1$ and $\alpha_2$ are two power-law indices, and parameter $\omega$ describes the sharpness of the break. The larger the value of the parameter $\omega$, the sharper the break in the SBPL function. Third, we perform a Monte Carlo simulation ($n = 1000$) based on the best-fitting SBPL function. The cumulative distribution of the simulated magnitude limits approximates the actual detection function of the instrument. Finally, the model fitting for this cumulative distribution is applied to find the intrinsic LF of the upper-limit sample. A similar simulation for the 18 GRBs in the detected sample is used to reconstruct the detected magnitude distribution. Combining the simulations of the limit and detected sub-samples, a simulated ``combined'' sample is then applied to constrain the LF obtained from ROTSE-III.

\section{Results}

\begin{figure*}[hp]
\hskip-0.3in
\includegraphics[angle=0,scale=0.7]{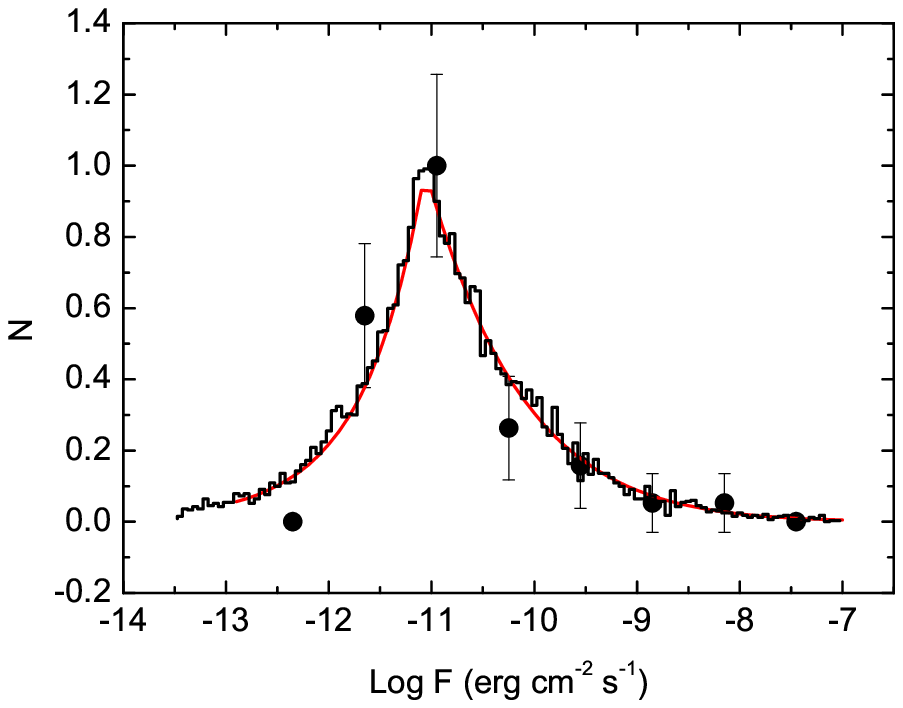}
\includegraphics[angle=0,scale=0.7]{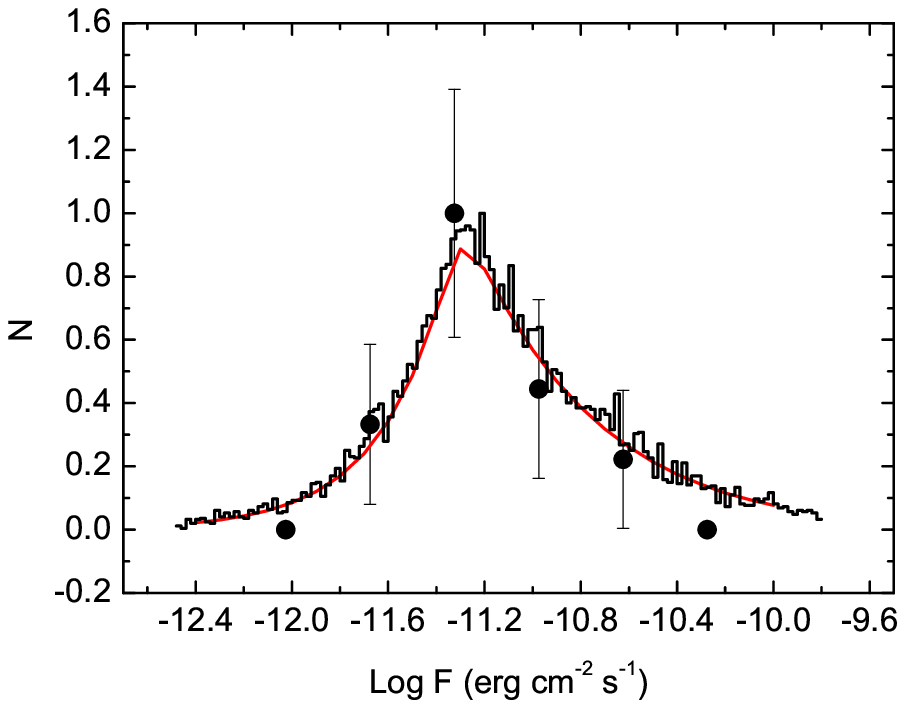}
\caption{The normalized histogram fitting (red solid curve) for the observed flux at 100s, after correcting for extinction in our Galaxy and the host galaxy, with a smooth broken power law (SBPL) based on the observations obtained with ROTSE-III (solid circles with errors) and 1000 Monte Carlo simulations (stepped line). The left panel shows the 40 GRBs in the upper-limit sample, and the right panel shows the 18 GRBs in the detected sample.}\label{fittings}
\end{figure*}

Figure \ref{fittings} shows the results of fitting the SBPL function to the simulation histograms for 40 limiting magnitudes (left panel) and 18 detected magnitudes (right panel) with red solid curves. The stepped lines are the Monte Carlo simulations in this figure. The best-fitting parameters, including the normalization factor $N_0$, the magnitude $f_b$ at the break point, the sharpness factor $\omega$, and the power-law indices $\alpha_1$ and $\alpha_2$ of SBPL, as described in Equation (\ref{SBPL}), are presented in Table \ref{smoothBPL}. The null hypothesis for the two groups, i.e., that the data from the observations obtained with ROTSE-III is from the same population as the simulations, is tested using a K-S test. The maximum distance between the two groups' cumulative probability functions is D-stat $= 0.28$, 0.19 with significance levels Prob=0.09, 0.12, respectively. This indicates that one cannot reject the null hypothesis (a common origin of the two samples) at the 5\% significance level, which gives confidence that the simulation based on the best fittings is appropriate in the case that the number of data points may not be large enough to construct the detection function of the instrument. The difference between the detection and limit sub-samples highlights the necessity to consider the detection function in the study of the GRB optical LF.
\vskip0.2in
\begin{table*}
\caption{Fit results and K-S test to the histogram of flux at 100s after trigger for the detected and upper-limit samples.}
\label{smoothBPL} \centering
%\begin{tabular}{|c|c|c|c|c|c|c|c|} \hline \hline
\begin{tabular}{c|ccccc|ccc} \hline \hline
parameter& N$_0$ & m$_b$ & $\alpha_1$ &  $\alpha_2$ & $\omega$& Prob & D-stat \\
\hline
detected &0.99& 11.31 &-20.76& 41.86& 5.84&0.09& 0.28\\
\hline
limit &0.99& 11.06 &-11.72 & 18.67 & 47.27&0.12 & 0.19\\
\hline\hline
\end{tabular}
\end{table*}

\begin{figure*}[hp]
\vskip0.2in
\hskip -0.1in
\includegraphics[angle=0,scale=0.5]{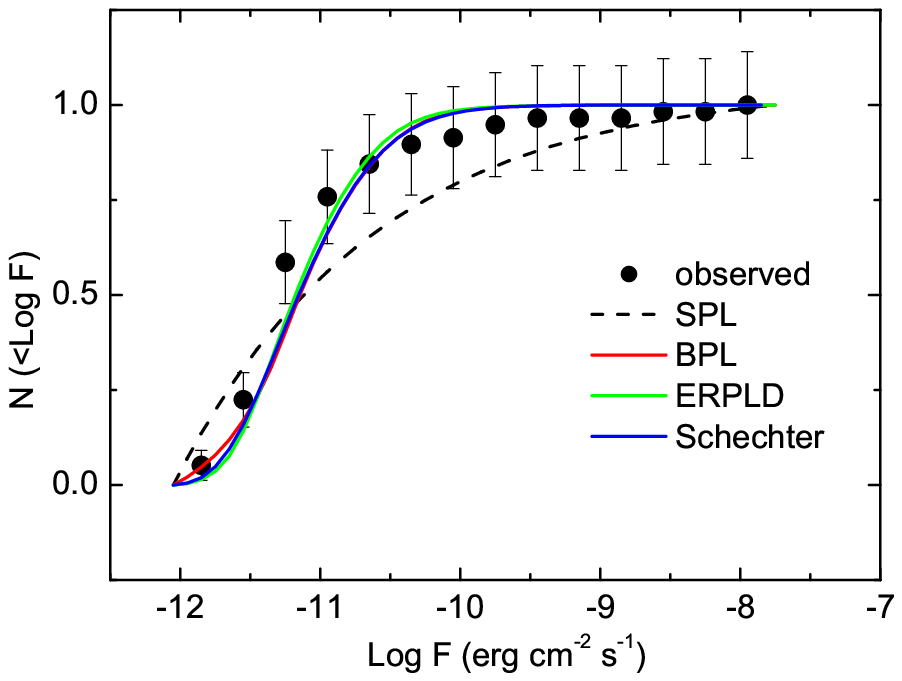}
\includegraphics[angle=0,scale=0.5]{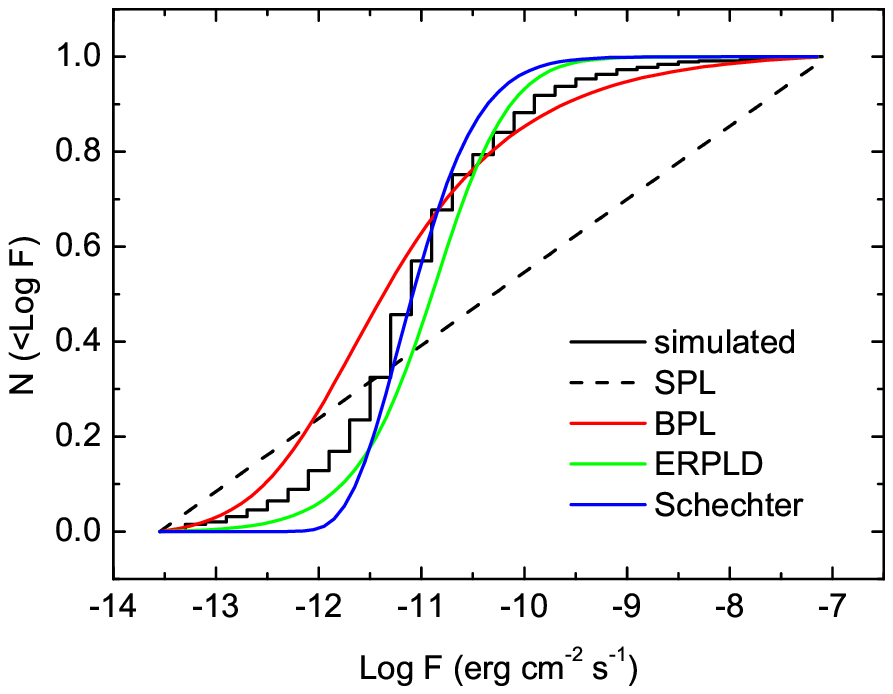}
\includegraphics[angle=0,scale=0.52]{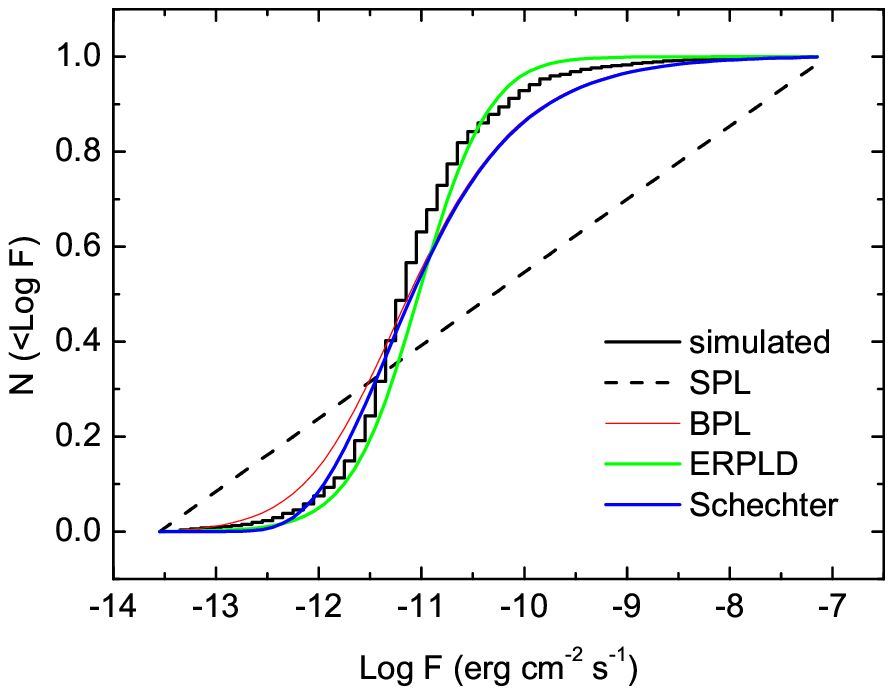}
\caption{The model fitting results for the cumulative distributions of 58 afterglows observed by ROTSE-III (left panel), simulated sensitivity function (middle panel) and the ``combined (simulated)'' sample including 1000 detection simulations and 1000 limit simulations. The solid circles with errors labeled in left panel as ``observed'' are the afterglows observed by ROTSE-III. The stepped lines are those from simulations. The type of LF is identified by color as described in the text.}\label{combine}
\end{figure*}

Figure \ref{combine} shows the cumulative distributions of the fluxes for the afterglows observed by ROTSE-III (solid circles with Poisson error bars in the left panel) and simulated results (stepped lines). The predictions of the flux distribution from the GRB formation rate based on SFR and different LFs are drawn with solid and dashed lines in this Figure. The optical LFs with different models (SPL, BPL, ERPLD, and Schechter function) are shown with different color in the Figure. The best fitting parameters for the models, as well as the results of the K-S test (D-stat, and significance level Prob), are presented in Table \ref{modelfit}. Considering that the significance level Prob is also a function of the number of data points, we calculate the value of Prob assuming a simulated number of 100 data points---not the number for the best fitting selection---in order to compare with the observed data. That is, the Prob values do not always correspond to confidence levels and we could not use this value as the standard to select the best fits.

From Table \ref{modelfit}, we find that the values of D-stat are smaller for the ``combined (simulated)'' sample (including 1000 detection simulations and 1000 upper-limit simulations) than those for the other two samples, including the ``obs'', data as observed by ROTSE-III, and the``limit (simulated)'', the simulated sensitivity function based on the modeled LF excluding the SPL model. This again implies that it is necessary to consider the sensitivity function of an instrument in studying the optical LF of GRBs.  For each data sample, the SPL (Eq. \ref{L1}) LF has the largest value of D-stat among all the LF models, 0.26 for the ``obs'' sample, 0.40 for the ``limited (simulated)'' sample, and 0.41 for the ``combined (simulated)'' sample. This can also be seen from Figure \ref{combine}, where the SPL model (dashed line) has the largest deviation from the ``obs'' data and simulated data (stepped line). We also find that the fits are insensitive to the values of $L_*$ for all the samples. For example, fits are insensitive to values of $L_{*}$ from 2 to 32 for the ``combined (simulated)'' sample. Excluding the SPL model, the values of D-stat for the ``combined (simulated)'' sample are in the range [0.07 0.08]. Furthermore, the BPL and Schechter function are also suitable models for the optical LF of GRBs at 100 s. But the values of D-stat are in the [0.10 0.12] range for ``obs'' data. For the ``limit (simulated)'' sample, the Schechter and ERPLD functions better describe the the sensitivity function of ROTSE with smaller values of D-stat (0.08 and 0.09), though the BPL function has one more parameter than the others.

\begin{table*}
\caption{The best-fit models of the cumulative distributions of flux for the afterglows observed by ROTSE-III (``obs'') and those for two simulations at 100s after trigger. One simulation is for the upper-limit sample (detection function). The other is for the ``combined (simulated)'' sample including 1000 detection simulations and 1000 upper-limit simulations.  The model functions are single power law (SPL), broken power law (BPL), exponential rise and power-law decay (ERPLD), and Schechter function.}
\label{modelfit} \centering
\begin{tabular}{|c|c|c|c|c|c|c|c|} \hline \hline
model& parameter& obs& limit (simulated) & combined (simulated)\\
\hline
\multirow{4}{*}{SPL}& $\alpha_L$& -1.3 & -1.0 & -1.0\\
\cline{2-5}
&$L_{*}$\dag(10$^{46}$ erg s$^{-1}$)& 25 & 34 &32  \\
\cline{2-5}
&D-stat& 0.26 & 0.40& 0.41 \\
\cline{2-5}
&Prob& 1.4 $\times 10^{-3}$ &1.1$\times 10^{-9}$& 2.7$\times 10^{-13}$ \\
\hline
\multirow{4}{*}{BPL}&$\alpha_{L1}$ &26.0& 1.4 & 1.6\\
\cline{2-5}
&$\alpha_{L2}$& 0.05 & 0.01& 0.01 \\
\cline{2-5}
&$L_{*}$(10$^{46}$ erg s$^{-1}$)& 14  & 0.9& 3.4 \\
\cline{2-5}
&D-stat& 0.11 & 0.12 & 0.07 \\
\cline{2-5}
&Prob& 0.52 & 0.06& 0.20 \\
\hline
\multirow{3}{*}{ERPLD}&$\alpha_L$ &4.9 & 0.01 & 0.4 \\
\cline{2-5}
&$L_{*}$(10$^{46}$ erg s$^{-1}$)& 1& 17 & 8\\
\cline{2-5}
&D-stat& 0.12& 0.09& 0.08 \\
\cline{2-5}
&Prob& 0.72 & 0.05& 0.25 \\
\hline
\multirow{3}{*}{Schechter}&$\alpha_L$ &-4.6  &-3.2& -1.6\\
\cline{2-5}
&$L_{*}$ (10$^{46}$ erg s$^{-1}$)& 20 & 13 & 2\\
\cline{2-5}
&D-stat& 0.10 &0.08 & 0.07 \\
\cline{2-5}
&Prob& 0.63 & 0.05& 0.15 \\
\hline\hline
\end{tabular}
\begin{tablenotes}
  \item[\dag]\dag The values from these best fits are not all strongly constrained, e.g., the fits are insensitive to values of $L_{*}$ from 2 to 32 for the ``combined (simulated)'' sample.
  \end{tablenotes}
\end{table*}

\section{Discussion and Conclusions\label{sec:conc}}
We construct the optical LFs of GRBs at 100s after the burst onset and study their functional form. The sensitivity function of the instrument is carefully considered with simulations, and we find it is necessary to take it into account for the study of the LFs of GRBs. We have found that an ERPLD, BPL or Schechter function is suitable model for the optical LF of GRBs observed by ROTSE III at 100s. An SPL functional form is excluded as the optical LF based on our GRB sample with high confidence.

We interpret the parameter $k$ in our sample as the ratio of GRBs detected by ROTSE-III in the field-of-view of \Swift, to all the bursts happening throughout the sky during the \Swift\ observation time. It is, however, difficult to determine the value of $k$; in the particular case of our study, the K-S test helps eliminate this parameter when finding the best fits, by normalizing the cumulative flux distribution of GRBs. An internal shock could produce the emission at the prompt phase and an external shock (reverse shock/forward shock) is thought to be a good candidate for the emission in the afterglow phase. However, the physical reason for the curved LF remains unclear. Kann et al. (2010) interpreted the three-Gaussian luminosity distribution as the existence of three ``classes'' of GRBs. It might be possible that the emission at 100s in our work originates from internal processes since they are earlier than those from the afterglow phase.

\begin{figure*}[hp]
\vskip0.2in
\hskip -0.1in
\includegraphics[angle=0,scale=0.8]{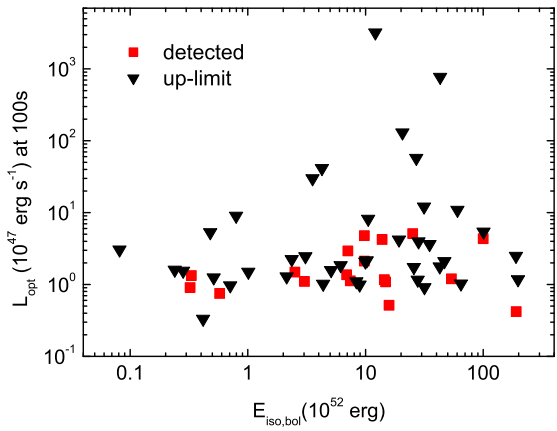}
\caption{The plot of optical luminosity $L_{\rm opt}$ at 100s after triggering versus the prompt isotropic bolometric energy $E_{\rm iso, bol}$ for 58 GRBs reported by ROTSE-III. The black downward pointing triangles are the upper-limit reports, and red squares are optical detections for 18 GRBs.}\label{EisoLo}
\end{figure*}

The optical luminosity was found to increase with increasing prompt energy release (Nysewander et al. 2009, Kann et al. 2010), similarly to the X-ray luminosity (e.g., Kouveliotou et al. 2004; Liang \& Zhang 2006; Amati et al. 2007; Gehrels et al. 2008). The plot of the optical luminosity $L_{\rm opt}$ at 100s after the burst onset versus the isotropic energy $E_{\rm iso,bol}$ radiated during the prompt phase of our sample is studied here to compare with previous work. There does not seem to be any trend of luminosity $L_{\rm opt}$ to energy $E_{\rm iso,bol}$ in our sample, as shown in Figure \ref{EisoLo}. The redshifts of GRBs in our sample were taken from Jochen Greiner's Table\footnote{http://www.mpe.mpg.de/$\sim$jcg/grbgen.html}. For GRB 110726A, the redshift of 1.036 is based on the only detected absorption line, whereas the upper limit of 2.7 is based on the non-detection of Lyman alpha; we adopt $z=1.036$ as its redshift. For the bursts without redshift measurements (21 GRBs in our sample), we assume redshifts of $z=2$ for the calculation of the luminosity distance of the bursts since the mean redshift of Swift GRBs has been shown to be close to 2 (e.g., Fynbo et al. 2009). The isotropic bolometric energies $E_{\rm iso, bol}$ released during the prompt phase of some GRBs in our sample have been calculated by Kann et al. (2010, 2011). For the bursts not included in the work of Kann et al. (2010, 2011), we calculated the values $E_{\rm iso, bol}$ based on Butler's analysis\footnote{http://butler.lab.asu.edu/swift/} (Butler et al. 2007) and the GCN report\footnote{http://gcn.gsfc.nasa.gov/gcn/}.

Rapid follow-up observations in the optical are critical to understand the physical processes of GRBs. There are quite a few small robotic telescopes, in addition to ROTSE-III, that have been built and installed around the world in order to rapidly search for GRB optical counterparts, e.g., GROCSE (Park et al. 1997), TAROT (Klotz et al. 2009), SkyNet\footnote{http://skynet.unc.edu/}, WIDGET (Urata et al. 2011),
MASTER\footnote{http://observ.pereplet.ru/}, Pi of the Sky (Burd et al. 2005), RAPTOR (Vestrand et al. 2002), REM (Zerbi \& Rem Team 2001), and Watcher (Ferrero et al. 2010). With their large Field of View (FOV) and fast slewing abilities, these telescopes offer promise to capture large samples of optical counterparts to accurately constrain the GRB LF at the earliest epoch after burst onset.
\vskip-0.2in
\acknowledgments
We thank the anonymous referee for his/her very constructive suggestions. We thank Z. G. Dai, B. Zhang, and L. P. Xin for helpful discussions. This work was supported by the National Basic Research Program of China (973 Program, grant No. 2014CB845800, and 2013CB834900) and the National Natural Science Foundation of China (Grant Nos. 11103026, 11322328 and 11025313). XFW acknowledges support by the One-Hundred-Talents Program, the Youth Innovation Promotion Association, and the Strategic Priority Research Program ``The Emergence of Cosmological Structures'' (Grant No. XDB09000000) of the Chinese Academy of Sciences. The research of JCW is supported in part by NSF Grant  AST-1109801.

\end{document}